\documentclass[10pt]{iopart}

\usepackage{graphicx}
\usepackage{xspace}
\usepackage{epstopdf}
\usepackage{textcomp}
\usepackage{booktabs}
\usepackage[pdftex]{}

\begin{document}

%\graphicspath{{Submission/}} % remove for submission

\title[]{Electroluminescence and current-voltage measurements of single (In,Ga)N/GaN nanowire light-emitting diodes in the nanowire ensemble}

\author{D van Treeck$^1$, J Ledig$^{2,3}$, G Scholz$^2$, J L\"{a}hnemann$^1$, M Musolino$^1$, A Tahraoui$^1$, O Brandt$^1$, A Waag$^2$, H Riechert$^1$ and L Geelhaar$^1$}

\address{$^1$ Paul-Drude-Institut f\"{u}r Festk\"{o}rperelektronik,
Hausvogteiplatz 5--7, D-10117 Berlin, Germany}
\address{$^2$ Institut f\"{u}r Halbleitertechnik, TU Braunschweig, Hans-Sommer-Str. 66, 38106 Braunschweig, Germany}
\address{$^3$ Physikalisch-Technische Bundesanstalt, Bundesallee 100, 38116 Braunschweig, Germany}

\ead{treeck@pdi-berlin.de}

\begin{abstract}
We present the combined analysis of the electroluminescence (EL) as well as the current-voltage (I--V) behavior of single, freestanding (In,Ga)N/GaN nanowire (NW) light-emitting diodes (LEDs) in an unprocessed, self-assembled ensemble grown by molecular beam epitaxy. The data were acquired in a scanning electron microscope equipped with a micromanipulator and a luminescence detection system. Single NW spectra consist of emission lines originating from different quantum wells, and the width of the spectra increases with decreasing peak emission energy. The corresponding I--V characteristics are described well by the modified Shockley equation. The key advantage of this measurement approach is the possibility to correlate the EL intensity of a single NW LED with the actual current density in this NW. This way, the external quantum efficiency (EQE) can be investigated as a function of the current in a single NW LED. The comparison of the EQE characteristic of single NWs and the ensemble device allows a quite accurate determination of the actual number of emitting NWs in the working ensemble LED and the respective current densities in its individual NWs. This information is decisive for a meaningful and comprehensive characterization of a NW ensemble device, rendering the measurement approach employed here a very powerful analysis tool.
\end{abstract}

\maketitle
\ioptwocol

%=====================================================================
\section{Introduction}
%=====================================================================
Group-III nitride nanowire (NW) ensembles have been employed for a wide range of applications, especially optoelectronic devices \cite{Zhao2015i}.
Considering the analysis of these devices, most studies focus mainly on the characterization of the NW ensemble properties. However, the properties of single NWs in the ensemble may differ considerably from the mean value measured for the ensemble.
For instance, in the particular case of light-emitting diodes (LEDs) based on self-assembled NW ensembles, the emission wavelengths of single NWs were found to vary substantially from wire to wire \cite{Kikuchi_2006,Lahnemann2011a,Bavencove_nt_2011,Limbach2012a}.  
Hence, in order to better understand the final behavior of NW ensemble devices, a careful analysis of the individual properties of the single NWs in the ensemble is mandatory.

In order to characterize NW ensemble LEDs in depth, a combined analysis of the emission and transport behavior of single NWs under electrical carrier injection is needed.
One way to do such a combined analysis is to remove the nanowires from the ensemble, disperse them on a substrate and contact them using lithographic methods.
However, the contact properties between the NW and the substrate, which might influence the overall performance of the single NWs in the working ensemble, can not be studied by investigating dispersed NWs. 
To overcome the disadvantages of the dispersion approach, single NWs can be directly contacted with a probe tip installed in a scanning electron microscope (SEM). For instance, using this technique, Lee \textit{et al.} studied the current-voltage (I-V) characteristics of single GaN based NW LEDs in the ensemble, however, they were not able to measure the respective electroluminescence (EL) signal \cite{Lee2011a}.  Yet another approach was implemented by Bavencove \textit{et al.} and Musolino \textit{et al} \cite{Bavencove_nt_2011,Musolino2017}. They did not contact single (In,Ga)N/GaN NWs with a probe tip, but detected diffraction-limited EL spots in the working ensemble device using a confocal microscope. However, with this approach they could not measure the currents in the investigated NWs.

Here, we present simultaneous measurements of the I--V behavior as well as the EL of single, freestanding nanowire LEDs in a self-assembled NW ensemble. To this end, as-grown NWs are contacted without any further processing with a probe tip installed in a SEM, which is equipped with a parabolic mirror in order to collect light emitted from the sample and couple it into a spectrograph. This method was already applied to investigate local electro-optical properties of much larger \textmu{}-LEDs and is now applied to nanostructures \cite{Ledig2014,Albert2015a,Ledig2016,Hartmann2017}.
Investigating in detail the spectral shape of the EL of single NWs, we find that different emission lines occur in the single NW spectra and that their width increases linearly with the peak emission wavelength. Furthermore, analyzing the I--V data of various NWs, we determine the series resistances as well as the threshold voltages of the single NWs. Finally, we analyze the dependence of the external quantum efficiency (EQE) on the current in single NW LEDs. By comparing the trends for single and ensemble measurements, we estimate the number of active NWs in the ensemble LED. This information allows us to determine fairly accurately the current density in the NWs in the working ensemble LED, which is an important parameter for device analysis.

%=====================================================================
\section{Experimental details}
%=====================================================================

\begin{figure*}[t!]
\begin{center}
\includegraphics*[width=16 cm]{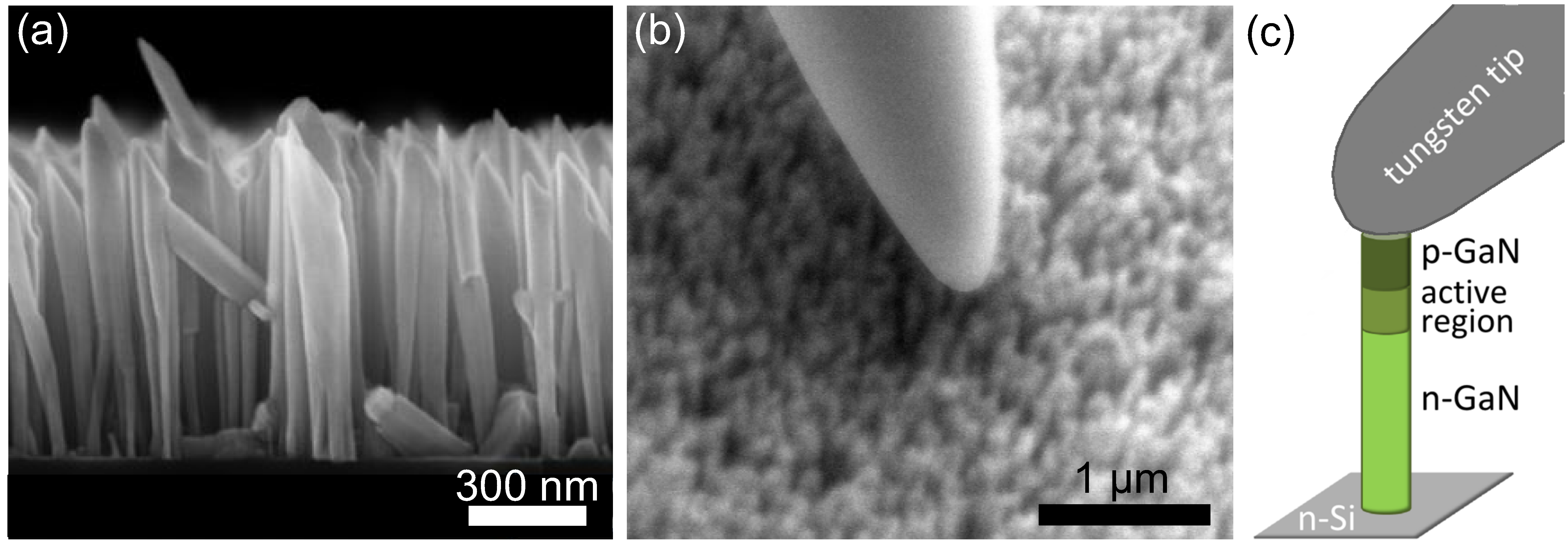}
\end{center} 
\caption[]{(a) Cross sectional micrograph of the investigated self-assembled GaN NW LEDs on Si acquired in the field-emission SEM. (b) Bird's eye view micrograph of a probe tip contacting single NWs taken in the thermionic emission SEM with a magnification at the edge of the resolution. (c) Sketch of a tungsten probe tip contacting a NW LED.}
\label{fig1}
\end{figure*}

The NW LED structures investigated in this work were grown by means of self-assembly processes with molecular beam epitaxy (MBE) on an n-doped Si(111) substrate. They consist of an intrinsic multiple quantum well structure grown on a Si doped n-GaN base of about 600\,nm length. The active region is composed of four (In,Ga)N insertions with an In content of (20$\pm$10)\% and a thickness of (3$\pm$1)\,nm. The insertions are separated by GaN barriers of 8\,nm thickness, where a 13\,nm wide, Mg-doped Al$_{0.15}$Ga$_{0.85}$N electron blocking layer follows the last insertion. On top of the active region a GaN:Mg cap of about 120\,nm length forms the p-type region. This results in NWs with a mean length of about 800\,nm and a mean diameter of around 100\,nm. Figure\,\ref{fig1}\,(a) shows a micrograph of the as-grown NW ensemble acquired in a Hitachi S-4800 field-emission SEM. 

%=====================================================================
%Fig.1
%=====================================================================

The NW ensemble LED we use as a reference for the EL and I--V characterizations in this study was processed from the same sample. The NW ensemble was planarized by spin coating using a solution of hydrogen silsesquioxane (HSQ). Subsequently, the upper 70 nm of the p-type segments of the NWs were uncovered by dry etching with CHF$_3$ and a 120-nm-thick indium tin oxide (ITO) layer was sputtered onto the NW tips. Finally, Ti/Au bonding pads and an Al/Au n-type contact were deposited on the top contacts and on the back side of the Si substrate, respectively.
A more detailed description of the growth and processing procedure as well as the EL and I--V characteristics of the NW ensemble LED can be found in Ref.\,\cite{Musolino2014}.

The EL and I--V measurements on single NW LEDs were carried out in a Zeiss DSM962 SEM system with an Everhart-Thornley SE detector. The resolution of the SEM system was optimized by means of reducing the probe energy and thus the scattering volume inside the NWs. However, in contrast to a field-emission SEM, the resolution is limited by the energy spread of the thermionic electron source resulting in chromatic aberrations. In order to contact nanostructures, the SEM system is equipped with a Kleindiek MM3A-EM micromanipulator which provides a metal tip of a nominal tip radius of 100\,nm [tungsten tip from GGB Industries Inc., visible in Figure\,\ref{fig1}\,(b)]. The probe tip as well as the sample holder are connected to a Keithley SMU 2635 source meter to apply and record the respective currents or voltages while the electron probe of the SEM is blanked. In an Oxford Instruments MonoCL2 the parabolic mirror above the sample holder collects light emitted from the sample and guides it out of the SEM chamber, through a monochromator to an Andor iDus 420 BV CCD camera. 
A slit width of 500\,\textmu{}m was chosen which corresponds, with respect to the grating ruling density of 150 lines per mm, to a resolution of about 7\,nm in the EL spectra. The obtained EL spectra were corrected taking into account the spectral responsivity of the system. 
An automatized measurement environment of this setup enabled a fast sequence of spectral and electrical operation points of contacted single NW LEDs in the ensemble while stepwise sweeping electrical injection. The EL and I--V measurements of all measurement positions shown in the manuscript (except position E) were acquired by sweeping the current with steps of 5\;nA. Only for measurement position E, we sweeped the voltage instead of the current with steps of 200\,mV. 

For this study, the EL and I--V measurements were performed at room temperature for various measurement positions on the sample, where one position corresponds to a single or a few contacted NW LEDs below the tungsten probe.
Using the SEM live mode, the piezo element-driven probe tip was approached very slowly to the NWs. The distance between probe tip and NW tips could be well estimated by comparing the shadow of the tungsten probe on the sample surface in the SEM live image with its actual position. When the shadow and the probe tip came together, the probe slightly changed its moving direction, thus indicating the contact to the sample [see Figure\,\ref{fig1}\,(b)]. The probe tip was then slowly lifted while applying voltage to monitor the electrical contact between NW and tip. With this procedure, it was possible to minimize the pressure while contacting the NWs. The spring force which is applied to the NWs in vertical direction once the probe is moved downwards, mainly results from the bending of the tungsten wire and can be estimated to be in the range of several tens of nN. Figure\,\ref{fig1}\,(c) shows a sketch of a single contacted NW LED.

%=====================================================================
\section{Results and discussion}

%=====================================================================
%=====================================================================
%Fig.2
\begin{figure*}[t!]
\begin{center}
\includegraphics*[width=10 cm,trim= 90 45 100 10]{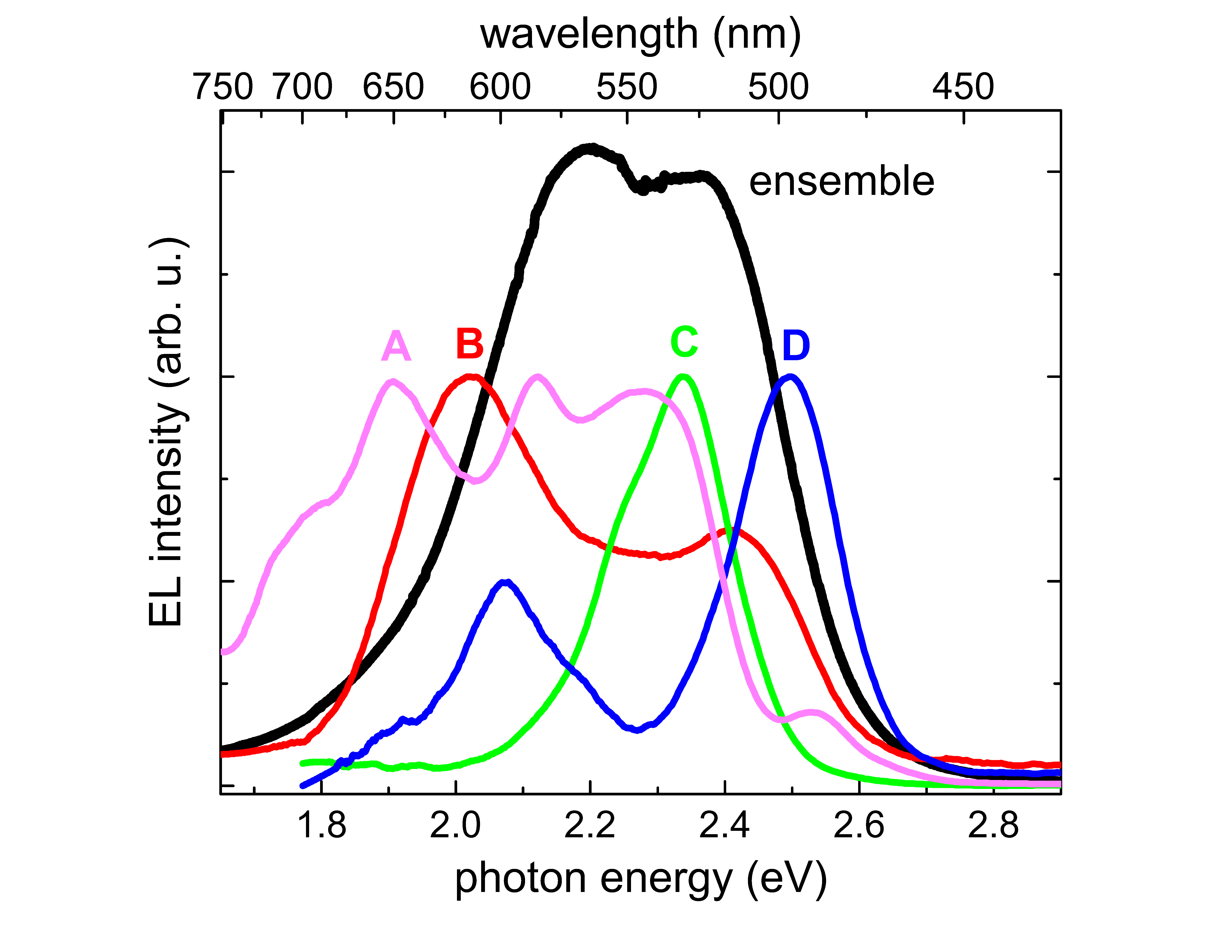}
\end{center} 
\caption[]{The colored spectra (normalized) represent the EL measured with the probe tip at four different positions on the unprocessed NW ensemble for a driving current of 100\,nA. The black spectrum is the EL of the processed ensemble LED \cite{Musolino2014}.}
\label{fig2}
\end{figure*}
%=====================================================================

Figure\,\ref{fig2} shows a representative collection of EL spectra obtained for four different measurement positions on the as-grown sample and the EL spectrum of the NW ensemble device processed from another piece of the same wafer. 
Comparing the spectra of the four measurement points, they show very different emission characteristics in terms of the number of emission bands, their emission energy and relative intensity.
For instance, spectrum C has one defined emission band with a broader tail towards lower energies, whereas for the other spectra at least two emission bands are visible. The emission bands can be found at various emission energies in a range from 1.77\,eV up to 2.53\,eV. 
The relative intensities differ from emission band to emission band and do not show a specific pattern. Nevertheless, the emission bands at lower energies are usually broader than the ones at higher photon energies. 
In general, it was not possible to draw a reasonable comparison between the absolute intensities of the different measurements, since the shadowing of the probe tip and the different measurement position with respect to the collection mirror is expected to have a strong influence.
Moreover, it should be noted that whenever the tungsten tip was in contact with the sample and a current (voltage) was applied, an EL signal could be detected. This is consistent with the conclusions of a previous investigation on similar samples that most of the NWs are well contacted to the substrate \cite{Limbach2012a}.

The comparison of the single NW measurements to the spectrum of the NW ensemble LED clearly suggests that the specific shape of the EL characteristic of the ensemble results from the superposition of the contributions of single NW LEDs with a substantial spread in emission properties. This finding is in agreement with the results of CL and EL studies on single NWs with a similar structure \cite{Lahnemann2011a,Bavencove_nt_2011,Musolino2017}.  
A more detailed discussion of the ensemble luminescence of the same NW ensemble can be found elsewhere \cite{Musolino2014}.

The resolution limit of the SEM and the restricted angle between sample and probe tip in combination with the high NW number density (5$\times$10$^9$\,cm$^{-2}$) of the investigated NW ensemble did not allow an identification of the number of contacted NWs during the various measurements by SEM. Hence, it is very likely that for the spectra A, B, and D in Figure\,\ref{fig2} more than one NW is contacted. A larger number of contacted NWs would result in a larger number of emission bands in the EL spectrum. 
Indeed, there are indications that the emission bands in the individual spectra of Figure\,\ref{fig2} originate from different single NWs.
For measurements where the probe tip is slightly moved sideways across the tips of the emerging NWs, one emission band after the other vanishes in the live-monitored EL spectrum and/or eventually new bands appear at different emission energies. In the supporting information we discuss in detail an example that shows the disappearance of an EL band due to a movement of the probe tip and the respective change in the I--V characteristics. 
Within a set of more than 20 measurement positions, spectrum C is representative for those spectra where only one major, slightly asymmetric emission band was visible. A similar band profile of the EL spectra of single NWs has already been reported in the literature  \cite{Bavencove_nt_2011,Musolino2017}. Hence, we assume that spectrum C shows the luminescence of a single nanowire LED.

%=====================================================================
%Fig.3
\begin{figure*}[t!]
\begin{center}
\includegraphics*[width=17 cm]{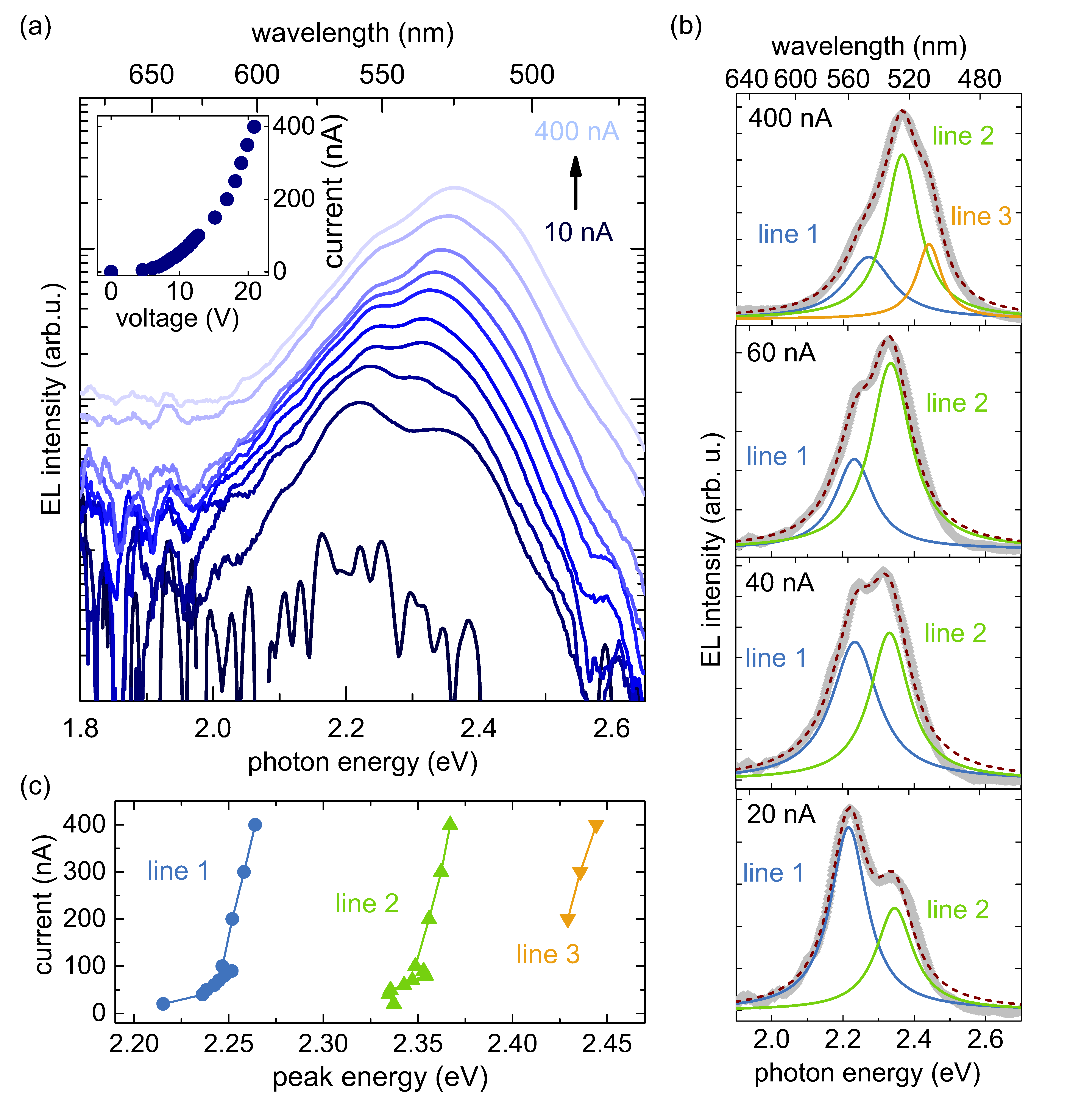}
\end{center} 
\caption[]{(a) EL spectra of NW C for currents in the range 10\,--\,400\,nA. The inset shows the corresponding current-voltage characteristic. (b)  The different line profiles in the spectra for driving currents of 20, 40, 60, and 400\,nA (data plotted as gray lines) were fitted by Lorentzian curves. The cumulative fit curves are displayed as purple dashed lines. (c) The graph shows the peak energy extracted from the fits of the different line profiles in (a) for driving currents from 20\,--\,400\,nA. For clarity the x-axis with the driving current is plotted vertically. Moreover, the lines connecting the different data points for each emission line are a guide to the eye.}
\label{fig3}
\end{figure*}
%=====================================================================

In order to better understand the emission behavior of single NWs, we now investigate the evolution of EL spectrum C with increasing driving current from 10\,to\,400\,nA, as illustrated in Figure\,\ref{fig3}\,(a). The first clear EL signal is obtained for a driving current of about 20\,nA. Assuming that at point C, we contact a single NW LED with a diameter of 100\,nm (mean value for the ensemble), the current range from 10\,to\,400\,nA would correspond to current densities in the single NW of about 130\,A/cm$^2$ to 5\,kA/cm$^2$, although it should be noted that it is not clear whether the current density is homogeneous across the NW. The respective I--V characteristic (inset) shows a clear diode behavior and hence indicates a stable contact between probe tip and NW. 
With increasing current two distinct emission lines around 2.24 and 2.34\,eV can clearly be distinguished in the spectra and for currents higher than 100\,nA, an additional shoulder becomes visible at about 2.43\,eV. 
Figure\,\ref{fig3}\,(b) shows that the EL spectra for the whole current range can be very well described by fitting the different emission lines with Lorentzian functions. The analysis demonstrates that for a driving current of 20\,nA the emission line at lower energy (line\,1) dominates, until the high energy line (line\,2) takes over at about 40\,nA. The shoulder that appears for currents higher than 100\,nA can be well modeled by a third emission line (line\,3), whose intensity increases with increasing current.
Analyzing the evolution of the peak energies of the emission lines 1, 2 and 3 for driving currents from 20\,to\,400\,nA as shown in Figure\,\ref{fig3}\,(c), we find that all emission lines experience a distinct blue-shift with increasing current. 
For clarity the x-axis with the driving current is plotted vertically in Figure\,\ref{fig3}\,(c).

The appearance of two main emission lines and their peculiar evolution with increasing current densities as shown for spectrum C is similar to what we recently observed analyzing single NW spots in top-view EL maps of a comparable processed NW ensemble LED \cite{Musolino2017}.
By modeling strain, electric field, and charge carrier density inside the active region of a single NW LED, it was found that the different emission lines in the spectra and their observed evolution with increasing injection current result from different emission energies and intensities of the four (In,Ga)N insertions. The low energy line was attributed to the EL of the first insertion (QW\,1) next to the n-type base of the NW, whereas the high energy line was linked to the superposition of the EL of the other three insertions. The variations in emission properties of single QWs were explained by different spontaneous and piezoelectric polarization fields within the different insertions, mainly caused by ionized donors and acceptors in the adjacent doped segments and by the non-uniform strain distribution along the active region, respectively.

Also in the case of NW C in Figure\,\ref{fig3}, the comparatively strong shift of emission line\,1 for currents up to 100\,nA suggests the presence of strong polarization fields within the contributing QW(s). 
Furthermore, the fluctuations of the peak energies of line\,2 for the current regime up to 100\,nA may be explained by the fact that line\,2 is a superposition of the emissions of several QWs with slightly different emission characteristics, as discussed above.
Note that for the low current regime, the contribution of emission line\,3, which is only well identifiable for currents higher than 200\,nA, might be hidden in line\,2. 
We note that the similarity between the line analysis in Ref.\,\cite{Musolino2017} and the data presented here corroborates our assumption that spectrum C corresponds to a single contacted NW.

%=====================================================================
%Fig.4
\begin{figure*}[t!]
\begin{center}
\includegraphics*[width=8 cm, trim= 5 5 0 0]{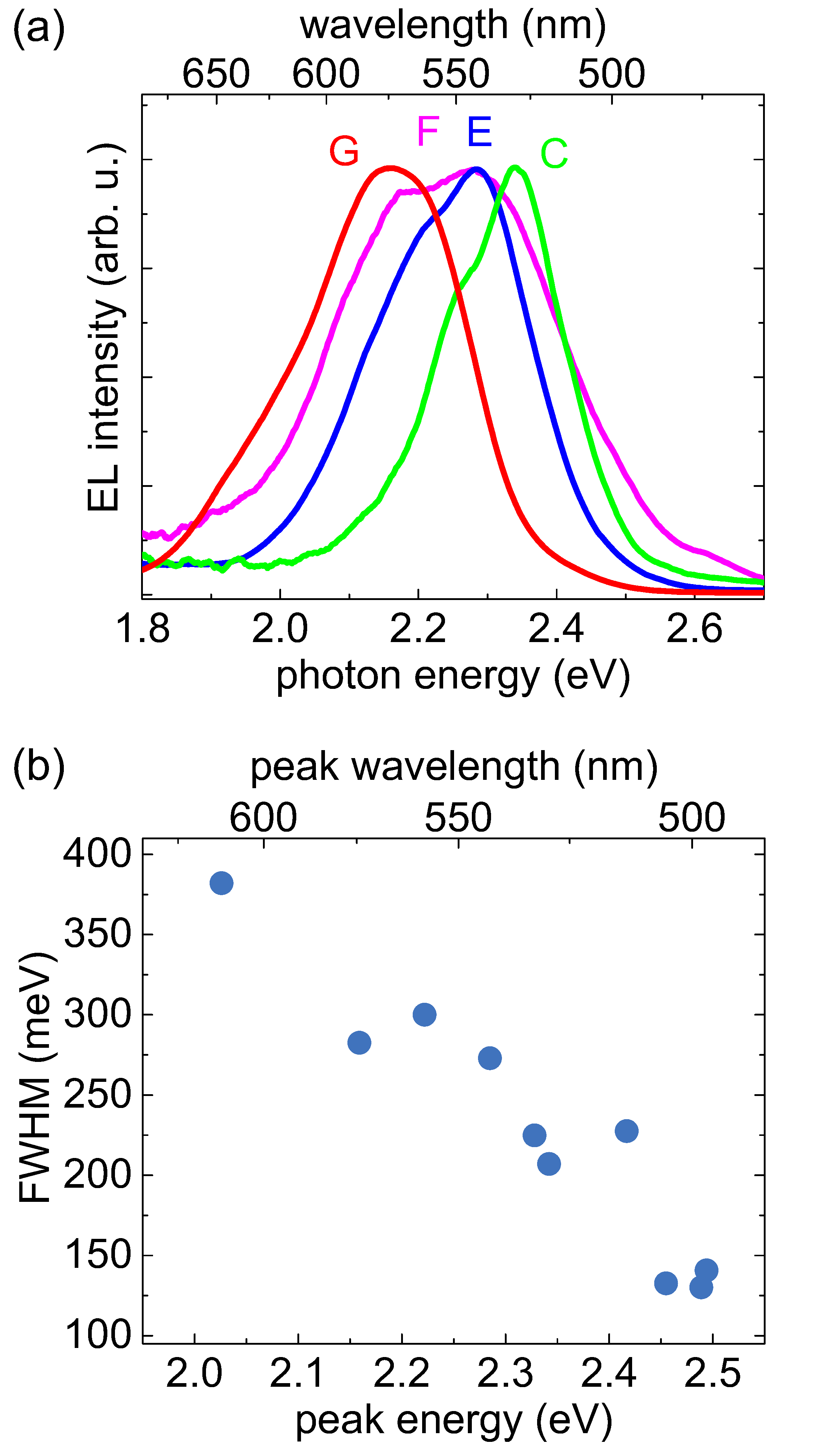}
\end{center} 
\caption[]{(a) Normalized EL spectra at various measurement points, including NW C for a driving current of 100\,nA ($\approx$\,1.3\,kA/cm$^2$). (b) FWHM of the EL spectra of single NW LEDs as a function of their peak emission energy.}
\label{fig4}
\end{figure*}
%=====================================================================

The previous analysis illustrates that contacting single NW LEDs in the ensemble is a powerful investigation tool to better understand the emission behavior of these structures under electrical carrier injection. 
In Figure\,\ref{fig4}, we now analyze and compare the emission properties of a set of different measurements on single NW LEDs.     
Figure\,\ref{fig4}\,(a) shows a comparison of the EL spectra of four different measurement points for a driving current of 100\,nA. The collection includes measurements of three single NW LEDs C, E and G emitting at different energies. The latter two were selected since they showed a similar emission behavior with increasing current as the previously analyzed NW C.
The fourth measurement point F exhibits a rather broad spectrum, so it can be assumed that at least two NWs are contacted by the probe tip. This point was chosen for comparison, and we will get back to it later.
Comparing the width of the single NW spectra C, E and G, one finds that it increases with decreasing emission energy. Figure\,\ref{fig4}\,(b) shows the full width at half maximum (FWHM) of ten EL spectra that could be attributed to different single NWs, including the values of C, E and G for a driving current of 100\,nA.
The width follows a linear decrease from about 380 to 130\,meV.
Such a trend may be explained by a higher local fluctuation of the material composition and a more inhomogeneous strain distribution within the insertions due to an increase in In content \cite{Qian2005,Wolz2012a}. 
Also for (In,Ga)N based core-shell NW structures grown by metalorganic chemical vapor deposition it was found that the FWHM increases with decreasing emission energy \cite{Qian2005}.

%=====================================================================
%Fig.5
\begin{figure*}[t!]
\begin{center}
\includegraphics*[width=8 cm, trim= 5 20 0 0]{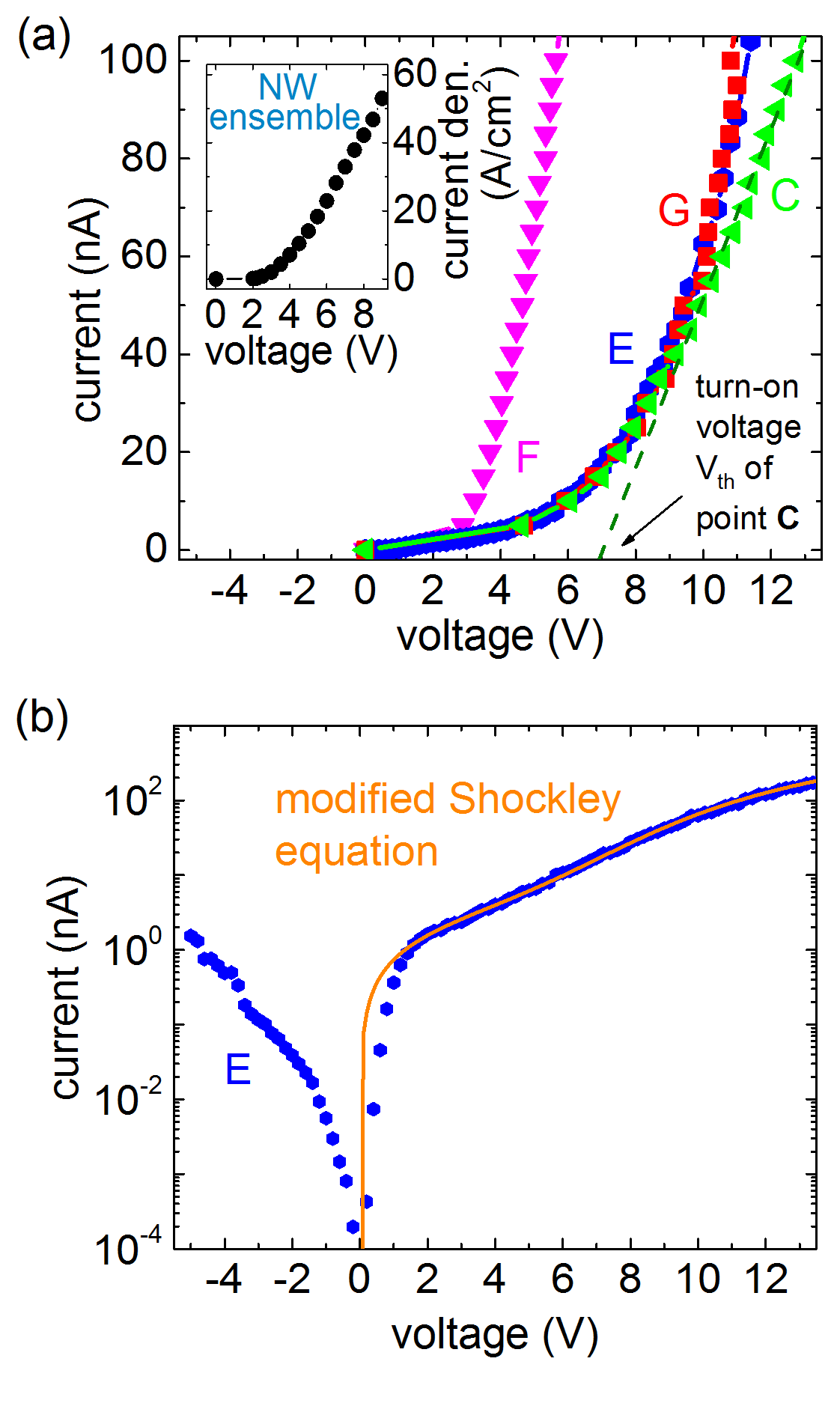}
\end{center} 
\caption[]{(a) I--V characteristic of points C, E, F, and G. The dashed line shows an example for a linear fit to the I--V curve of NW C for the current range from 60\,--\,100\,nA from which the series resistance $R_\mathrm{tot}$ and the threshold voltage $V_\mathrm{th}$ are derived. The inset depicts the I--V behavior of the ensemble. (b) Semi-log plot of the I--V curve (blue dots) of NW E and the respective fit (orange line) of the data, obtained by using Eq.\ref{shockley}.}
\label{fig5}
\end{figure*}
%=====================================================================

In order to better understand the charge carrier transport in single NW LEDs, in Figure\,\ref{fig5}\,(a) we analyze the I--V curves for the NWs C, E, and G, as well as for the bundle of NWs F for currents in the range 0\,--\,100\,nA. 
The I--V characteristic of the NW ensemble LED is displayed for comparison in the inset. All curves exhibit a clear diode behavior and for currents higher than about 60\,nA, the I--V curves show a linear behavior. In this region, the curves differ in slope and thus in series resistance. The series resistances $R_\mathrm{tot}$ of C, E, F and G, evaluated from a linear fit of the I--V curves for the range 60\,--\,100\,nA are shown in Table\,\ref{tab1}.

\begin{table}[b!]
\centering
\begin{tabular}{ccccc}
\br

LED(s) & & $R_\mathrm{tot}$ ($\Omega$) &   & $V_\mathrm{th}$ (V) \\
\mr
C   & &58\,$\times$\,10$^6$ &       & 7.0 \\
E	& &43\,$\times$\,10$^6$ &    & 7.3\\     
F	& &22\,$\times$\,10$^6$ &    & 3.5 \\
G	& &34\,$\times$\,10$^6$ &    & 7.8 \\
ensemble & &30 &     & 3.3 \\
\mr
\end{tabular}
\caption{\label{tab1}. Series resistance $R_\mathrm{tot}$ and threshold voltage $V_\mathrm{th}$ of the single NWs C, E, and G, as well as of the bundle of NWs F and the NW ensemble.}
\end{table}

In general, the main contributions which cause the high values of $R_\mathrm{tot}$ are the resistance of the n-GaN/n-Si interface $R_\mathrm{GaN/Si}$, the contact resistance $R_\mathrm{c}$ between p-GaN and the tungsten probe and the resistance of the NW LED $R_\mathrm{NW}$ itself. The contributions of the n-Si substrate and the measurement setup were determined to be in the range of a few Ohms and therefore can be neglected for our considerations. 
Unfortunately, it was not possible to separate the single contributions of $R_\mathrm{GaN/Si}$, $R_\mathrm{c}$ and $R_\mathrm{NW}$ in our measurements.
For comparison, we estimate the resistance of a single NW in the working NW ensemble LED by $R_\mathrm{tot}^\mathrm{est.}= A\, d_\mathrm{on} R_\mathrm{S}$ with the device area $A$, the series resistance $R_\mathrm{S}$ and the number density of emitting NWs $d_\mathrm{on}$. With values of 0.19\,mm$^2$, 30\,$\Omega$ and about 3\,$\times$\,10$^8$\,cm$^{-2}$ for the ensemble (see Ref.\,\cite{Musolino2014}), respectively, one obtains a value of 17 M$\Omega$, which is comparable to the values obtained for the single NW measurements (Table\,\ref{tab1}). 
Later in this report, we present an independent estimate of $d_\mathrm{on}$ that is higher by a factor of two, which results in an even better agreement between single NW and ensemble $R_\mathrm{tot}$.
Nevertheless, a value of $R_\mathrm{c}$ of several M$\Omega$ in the case of single unprocessed NW LEDs contacted by a probe tip cannot be excluded.
The comparatively small $R_\mathrm{tot}$ of point F can be explained by the fact that for this measurement point most certainly two NWs are contacted, resulting in half the series resistance as for the single NW measurements C, E, and G.

Another parameter that can be extracted from the I--V behavior is the threshold voltage $V_\mathrm{th}$, which is defined as the zero-crossing of the linear fit function of the respective I--V curves. As an example, the dashed line shows the linear fit for NW C. The threshold voltages for the different measurements are shown in Table\,\ref{tab1}.
These values are in agreement with the actual turn-on voltage for which the first EL signal was detected. The comparatively high threshold voltages of single NW measurements in comparison to the 3.3\,V of a NW ensemble LED with a processed ITO top-contact most likely result from a Schottky type contact at the p-GaN/tungsten interface.

To analyze the I--V behavior of a single NW in more detail, Figure\,\ref{fig5}\,(b) shows the I--V curve of NW E in semi-log scale. 
For positive voltages, the current increases continuously with increasing voltage and can be well described by the modified Shockley equation as it is commonly used in the literature \cite{Morkoc2008,Lee2011a,Mohammad2010,Suzue1996,Kim2002},
\begin{equation}{}
\label{shockley}
I= I_0 \left[\exp \left(\frac{e (V - I R_S)}{n~k_B T} \right)-1 \right]+  \frac{V - I R_S}{R_P}.
\end{equation}
From the fit (orange line), values for the saturation current $I_0$ and the ideality factor $n$, as well as the parallel resistance $R_\mathrm{P}$ and a series resistance $R_\mathrm{S}$ of the contacted NW LED or NW LEDs could be obtained, which are 80\,pA, 50, 1.5\,G$\Omega$ and 18\,M$\Omega$, respectively. 
For positive voltages <\,2\,V, the I--V behavior is dominated by the parallel leakage current defined by $R_\mathrm{P}$. The voltage regime 2\,--\,9\,V is mainly characterized by the diode term where the ideality factor $n$ determines the slope of the curve until $R_\mathrm{S}$ dominates the current for voltages >\,9\,V. With a value of 50, the ideality factor $n$ is far from unity. Also other studies of single NW diodes based on GaN obtained ideality factors of about 20\,--\,30 \cite{Lee2011a,Mohammad2010,Kim2002,Tchernycheva2014,Brubaker2013} or even 161 \cite{Lin2009}. Regarding the origin of these high ideality factors, it was found that in particular the contact to the p-type GaN, in our case the tungsten/p-GaN interface, may be responsible for ideality factors $\gg\,2$ \cite{Shah2003,Zhu2009}.
We emphasize that for the NW ensemble LED with optimized ITO top contact to the p-GaN NW tips the ideality factor is with 9.2 \cite{Musolino2014} much smaller than for the single NW measurement presented here. Thus, we conclude that the high ideality factor found here is caused by the contact between the tungsten tip and the p-GaN or, much less likely, the absence of the dielectric covering the nanowire sidewalls in the processed LED, but not related to processes inside the NWs themselves. The large difference in the ideality factor points to limitations of the single NW measurements for a detailed analysis of IV characteristics.

For negative voltages the I--V characteristic shows a rapidly increasing and rather high reverse leakage current, which cannot be described by the modified Shockley equation. 
At $-5$\,V the reverse current is with 1.5\,nA about one fourth of the forward current at 5\,V.  
For a more detailed discussion of the reverse leakage current we refer the reader to our previous study \cite{Musolino2016} where we present a comprehensive model that describes quantitatively the I--V characteristic of nanowire LEDs under reverse bias.

In contrast to investigating EL maps of the NW ensemble LED, one major advantage of our measurement approach is that one can correlate the EL intensity of a single NW LED with the actual current flowing through this NW. This allows a current-dependent analysis of the integrated EL intensity and the relative EQE of single NW LEDs in the ensemble. 
The relative EQE is defined by the quotient of the integrated EL and the respective driving current \cite{Morkoc2009}. The term "relative" takes into account that the total emitted intensity at the different measurement positions is unknown due to shadowing effects of the probe tip and different angles between measurement positions and collection mirror. Hence, no absolute values for the EQE could be determined.
Furthermore, for a given, externally imposed driving current, we can assume that the influence of any high contact resistance between probe tip and NW LED on the integrated EL and hence the relative EQE is negligible.

\begin{figure*}[t!]
\begin{center}
\includegraphics*[width=8 cm, trim= 9 5 10 0]{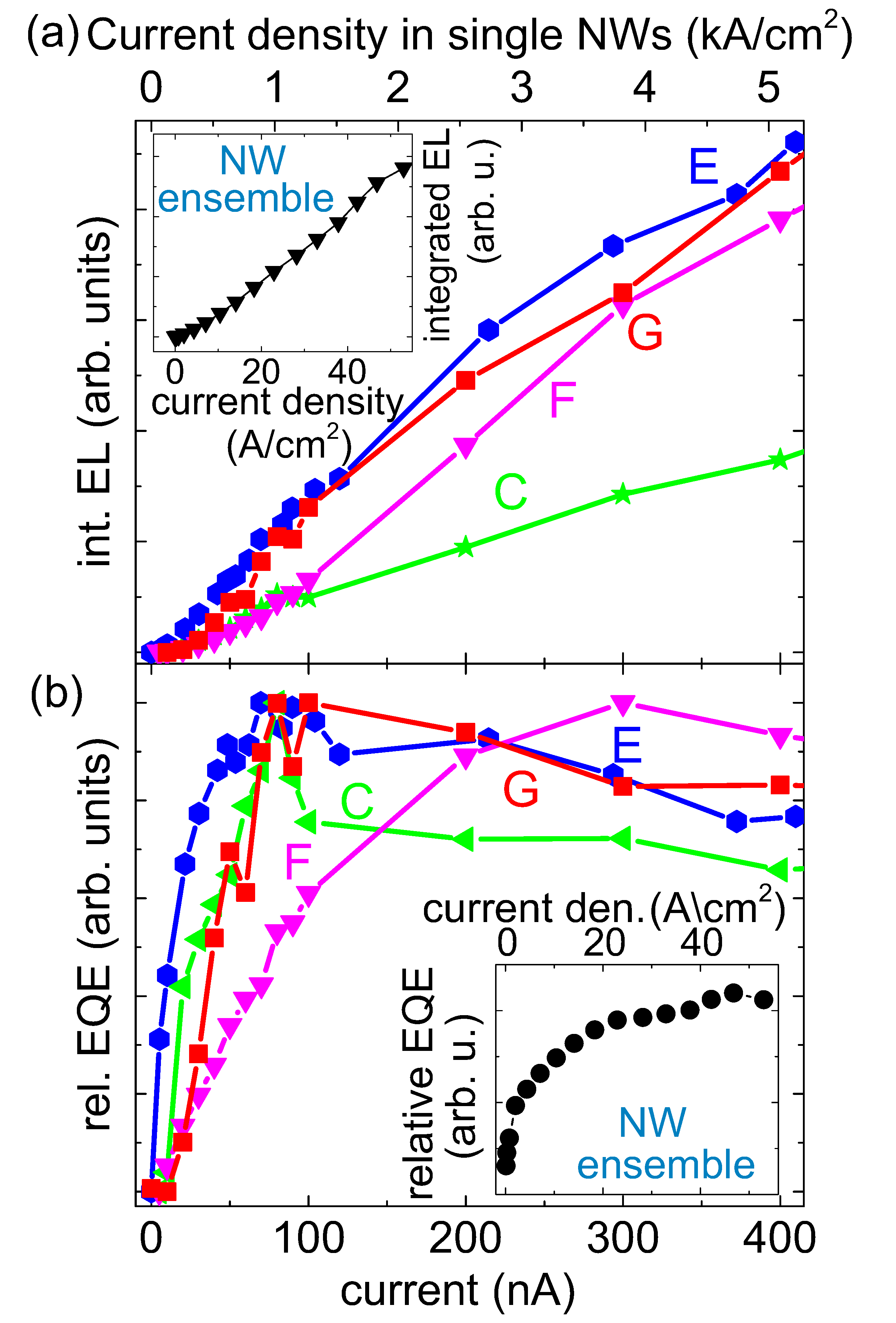}
\end{center} 
\caption[]{\textbf{(a)} Integrated EL and \textbf{(b)} normalized relative EQE of the measurement points C, E, F, and G as a function of current. The respective insets in the graphs (a) and (b) show the integrated EL and relative EQE of the NW ensemble LED.}
\label{fig6}
\end{figure*}

In Figure\,\ref{fig6} the integrated EL and the relative EQE are shown for the same collection of measurement points C, E, F, and G. The respective current densities in the single NW LEDs C, E, and G for this range are given at the top x-axis and were estimated using the extracted mean NW diameter of 100\,nm. These values are not valid for point F, since here more than one NW is contacted resulting in lower current densities. The integrated EL was obtained by integrating the single EL spectra for the different driving currents. 
However, as explained above, one has to be careful with directly comparing the absolute values of the integrated EL for the different measurement points. 
Nevertheless, the current dependence of the integrated EL should be meaningful. For most of the measurements a continuous linear increase of the integrated EL with the driving current was observed over the whole measurement range. A similar trend was also found for the ensemble LED, which is shown in the inset of Figure\,\ref{fig6}\,(a) \cite{Musolino2014}.

%=====================================================================
%Fig.6
%=====================================================================

The relative EQE of the different points plotted in Figure\,\ref{fig6}\,(b) shows a maximum for all the single NW measurements C, E, and G at a current value of around 80\,nA, which corresponds to a current density of about 1\,kA/cm$^2$. The maximum of point F only appears for a current of about 300\,nA. Once the maximum relative EQE is reached, the EQE slightly decreases but remains at a rather high level.
In general, the majority of the measurements where single NW LEDs were contacted, showed a relative EQE maximum for injection currents between 60 and 100\,nA. 
Interestingly, there seems to be a correlation between the appearance of the relative EQE maximum and the presence of strong piezoelectric fields within the QWs, as suggested by the pronounced spectral blue shift of the EL lines for currents below 100\,nA in Figure\,\ref{fig3}\,(c). 
Strong polarization fields cause a spatial separation of electrons and holes in the QWs which in turn leads to a reduced radiative recombination rate and hence a reduced relative EQE for small injection currents.
Due to the lack of temperature dependent measurements, it was not possible to draw any hard conclusions about the origin of the slight decrease of the relative EQE once the maximum was reached and its rather constant subsequent behavior for higher current densities. The absence of a significant relative EQE drop for NW LEDs for high current densities was also reported in several studies on NW ensemble LEDs \cite{Nguyen2012a,Zhao2016}.

The comparatively high current for which the relative EQE of point F peaks, can be explained by taking into account that for this point more than one NW is contacted. Hence, for the same driving current at point F, the mean current density in the single contributing NWs is lower. As a consequence, for an increasing number of contacted NWs, the maximum in the relative EQE only appears for higher driving currents, since the current $I^\mathrm{EQE_\mathrm{max}}_\mathrm{NW}$ for which the saturation of the relative EQE of the single NW LEDs sets in, is not yet reached. Considering a NW ensemble LED, this means that the device current density $J^\mathrm{EQE_\mathrm{max}}_\mathrm{device}$ (driving current divided by device area) for which the EQE of the device has its maximum, strongly depends on the number density of emitting NWs $d_\mathrm{on}$ in the ensemble and is given by the relation $J^\mathrm{EQE_\mathrm{max}}_\mathrm{device}=d_\mathrm{on}I^\mathrm{EQE_\mathrm{max}}_\mathrm{NW}$. Hence, the lower the number density of active single NW LEDs in the ensemble, the lower is the device current density for the maximum relative EQE and vice versa. It should be noted that this relation is only true if the NWs have similar contact resistances, resulting in a homogeneous current spreading in the ensemble device. 
The side-by-side comparison of ITO and Ni/Au top contacts in Ref.\,\cite{Musolino2014}  showed that with ITO a homogeneous p-type contact can be achieved throughout the whole device. This was not the case for the Ni/Au top contact, for which the contact resistance varied significantly across the sample. Considering the EQE characteristics of a device, such a pronounced variation leads to a rather high, strongly varying contact resistance resulting in very different currents in the single NWs and hence a slow rise of the ensemble EQE.

The relative EQE of the ensemble device with the ITO top contact fabricated from the same sample investigated in this study is shown in the inset of Figure\,\ref{fig6}\,(b). The graph depicts that the relative EQE saturates for a device current density $J^\mathrm{EQE_\mathrm{max}}_\mathrm{device}$ of about 47\,A/cm$^2$. Taking the latter value and the value for $I^\mathrm{EQE_\mathrm{max}}_\mathrm{NW}$ of 80\,nA we obtain from our measurements on single NW LEDs, one can estimate the actual number density of emitting NWs $d_\mathrm{on}$ in the ensemble, which is 6\,$\times$\,10$^8$\,cm$^{-2}$, using the above mentioned relation. With a number density of about 5\,$\times$\,10$^9$\,cm$^{-2}$ for the as-grown ensemble, this corresponds to about 12\% emitting NWs in the processed NW ensemble which contribute to the total EL emission. This is in agreement with the rough estimation obtained by analyzing the ensemble LED \cite{Musolino2014}, but the current procedure provides a more precise and reliable value. The fairly low fraction of active NWs can be attributed to the complex post-growth processing of NW LEDs, where, e.\,g., a homogeneous planarization of the NW ensemble and homogeneous ITO top-contact to the p-GaN tips of the NWs are crucial factors which strongly influence the number of emitting NWs \cite{Musolino2014,Limbach2012a}. Furthermore, the current densities in single NW LEDs in the working NW ensemble device can be estimated to be in the range from 20\,A/cm$^2$\,to\,1\,kA/cm$^2$ for device current densities of 0.9\,--\,47\,A/cm$^2$. We emphasize that without the single NW measurements introduced here, calculations of the actual current density per active NW in an ensemble device are limited in accuracy since it cannot be easily determined how many NWs actually participate in charge conduction.

%=====================================================================
\section{Summary and Conclusions}

%=====================================================================

In this study, we employed a specially equipped SEM to contact as-grown single (In,Ga)N/GaN NW LEDs with a probe tip and acquire simultaneously I--V characteristics and EL. Even though the NW number density of our sample is so high that direct imaging by the SEM does not reveal the number of contacted NWs, a careful analysis of electrical and EL data allows to distinguish with confidence between cases where single and multiple NWs are contacted. The emission energy of individual NWs varies, and the ensemble EL spectrum can be understood as a superposition of the individual spectra. The FWHM of the individual spectra decreases with increasing emission energy, which is consistent with previous reports \cite{Qian2005} and can be explained by compositional fluctuations as well as by inhomogeneous strain distribution within the insertions being more pronounced for higher In contents \cite{Qian2005,Wolz2012a}. The individual spectra consist of three emission lines whose intensity changes in a characteristic way with current. This phenomenon is caused by the N polarity of the NWs and the three-dimensional strain profile resulting from elastic relaxation at the free sidewall surfaces, as previously shown in Ref.\,\cite{Musolino2017}.

I--V curves acquired for single NW LEDs exhibit diode characteristics similar to ensemble measurements. However, the threshold voltage and ideality factor are significantly higher for the single NW experiments, likely because of a high contact resistance between tungsten probe tip and p-GaN NW top. Taking this effect into account, the two types of measurement are consistent. An important result is that the single NW analysis confirms the high leakage current under reverse bias found for the ensemble. This agreement implies that the leakage behavior is inherent to the as-grown NW structure and is not caused by deficiencies in processing. 
 
The key advantage of our measurement approach is the possibility to correlate the EL intensity of a single NW LED with the actual current density in this NW. Qualitatively, EL intensity and relative EQE increase with current as seen for the ensemble LED, and the latter trend exhibits a maximum followed by a slight decrease. However, the decisive difference is that for the ensemble measurement the fraction of NWs participating in charge transport and emitting EL can only be roughly estimated, essentially because the NWs in the dense ensemble cannot be optically resolved \cite{Musolino2014}. Furthermore, from the comparison of different processing protocols it is known that it is very challenging to obtain EL from a substantial fraction of NWs \cite{Limbach2012a}. In contrast, for the single NW measurements introduced here, the current density is known fairly well. By comparing the currents at which the maximum in relative EQE occurs for single NW and ensemble measurements, we could determine for the ensemble the fraction of active NWs emitting EL and the current density in them. The value of only 12\% active NWS we found, implies that there is still a large potential for the optimization of processing, which would lead to significant improvements of the overall performance of NW LEDs. More importantly, information about the actual current density in the semiconductor heterostructure is crucial for a meaningful assessment of NW ensemble devices, in particular in comparison with planar devices. This information is equally decisive for the comparison of device simulations as reported in Ref.\,\cite{Musolino2017} with experimental results for NW LEDs. Therefore, the measurement technique employed here is a very powerful analysis tool for the investigation of LEDs based on NW ensembles and provides new opportunities for their detailed study.

%=====================================================================
\ack % Acknowledgements
%=====================================================================

We are grateful to C. Herrmann, H.-P. Sch{\"{o}}nherr, and C. Stemmler for the maintenance of the MBE system, and A.-K. Bluhm for SE micrographs. Furthermore, we are thankful to Alexander Kuznetsov for a critical reading of the manuscript. Financial support by the European Commission (Project DEEPEN, FP7-NMP-2013-SMALL-7, Grant Agreement no. 604416) is gratefully acknowledged.

%=====================================================================
\section*{References}
%=====================================================================

\bibliographystyle{iopart-num}
\bibliography{manuscript}

\end{document}